\documentstyle[aps,preprint]{revtex}\tighten
\def\D{\mbox{D}}
\def\c{\mbox{curl}\,}
\begin{document}

\title{Cosmological magnetic fields}

\author{Roy Maartens\thanks{roy.maartens@port.ac.uk}
\thanks{Based on a plenary talk given at the International Conference on
Gravitation and Cosmology, January 2000, India}}
\address{
Relativity and Cosmology Group, School of Computer Science and
Mathematics,\\ Portsmouth University, Portsmouth~PO1~2EG, Britain}

\maketitle

\begin{abstract}

Magnetic fields are observed not only in stars, but in galaxies,
clusters, and even high redshift Lyman-$\alpha$ systems. In
principle, these fields could play an important role in structure
formation and also affect the anisotropies in the cosmic microwave
background radiation (CMB). The study of cosmological magnetic
fields aims not only to quantify these effects on large-scale
structure and the CMB, but also to answer one of the outstanding
puzzles of modern cosmology: when and how do magnetic fields
originate? They are either primordial, i.e. created before the
onset of structure formation, or they are generated during the
process of structure formation itself.
\end{abstract}

\[ \]

~\\

\section{Introduction}

Magnetic fields seem to be everywhere that we can look in the
universe, from our own sun out to high-redshift Lyman-$\alpha$
systems. The fields we observe (based on synchrotron radiation and
Faraday rotation) in galaxies and clusters have been amplified by
gravitational collapse and possibly also by dynamo mechanisms.
They are either primordial, i.e. originating in the early universe
and already present at the onset of structure formation, or they
are protogalactic, i.e. generated by battery mechanisms during the
initial stages of structure formation. One way to distinguish
these possibilities would be to detect or rule out the presence of
fields coherent on cosmological scales during recombination via
their imprint on the cosmic microwave background (CMB) radiation.
The new generation of CMB observations (especially the MAP and
Planck satellites) may be able to achieve this.

The origin, evolution and cosmological impact of magnetic fields
represent a fascinating challenge to theorists. I will discuss
some aspects of this challenge in the following sections.
(See~\cite{review} for some other recent reviews.) Some basic
facts from magnetohydrodynamics will be useful for the discussion.

Maxwell's equations are
\begin{equation}\label{1}
\nabla_{[\mu}F_{\nu\alpha]}=0\,,~~ \nabla_\nu F^{\mu\nu}=J^\mu\,,
\end{equation}
where $F_{\mu\nu}=\nabla_\mu A_\nu-\nabla_\nu A_\mu$ is the field
tensor, $A_\mu$ is the four-potential, and $J_\mu$ is the
four-current. The field tensor is observer-independent, while the
electric and magnetic fields depend on the observer's motion:
\begin{equation}\label{2}
E_\mu=F_{\mu\nu}u^\nu\,,~~B_\mu={\textstyle{1\over2}}
\varepsilon_{\mu\nu\alpha}F^{\nu\alpha}\,,
\end{equation}
where $u^\mu$ is the observer's four-velocity, and
$\varepsilon_{\mu\nu\alpha}$ is the covariant permutation tensor
in the observers' rest space.

Ohm's law is
\begin{equation}\label{3}
h_{\mu\nu}J^\nu=\sigma F_{\mu\nu}u^\nu\,,
\end{equation}
where $h_{\mu\nu}=g_{\mu\nu}+u_\mu u_\nu$ projects orthogonal to
$u^\mu$ (and $g_{\mu\nu}$ is the metric). For most of the history
of the universe, the conductivity $\sigma$ is extremely high. In
the magnetohydrodynamic limit, we have $\sigma\to\infty$ while the
current remains finite, so that $E_\mu\to0$. Thus the electric
field in the particle frame vanishes: $F_{\mu\nu}u^\nu=0$. In the
observer's frame, with four velocity $\tilde{u}^\mu=u^\mu+v^\mu$,
where $v^\mu$ is the relative velocity ($v_\mu u^\mu=0$) and we
neglect terms $O(v^2)$, the electric field is of course not zero,
but given by
\begin{equation}
\tilde{E}_\mu=-\varepsilon_{\mu\nu\alpha}v^\nu B^\alpha\,.
\end{equation}
In this limit, Maxwell's equations may be written as~\cite{tb,tm}:
\begin{eqnarray}
\D^\mu B_\mu &=& 0\,,\label{4}\\ \omega^\mu B_\mu & =&
-{\textstyle{1\over2}}J_\mu u^\mu\,, \label{5}\\ \c B_\mu &=&
h_{\mu\nu}J^\nu +\varepsilon_{\mu\nu\alpha}B^\nu \dot{u}^\alpha\,,
\label{6}\\h_{\mu\nu}\dot{B}^\nu &=& -{\textstyle{2\over3}}\Theta
B_\mu+\sigma_{\mu\nu}B^\nu+\varepsilon_{\mu\nu\alpha} B^\nu
\omega^\alpha\,,\label{7}
\end{eqnarray}
where $\D_\mu$ is the projected covariant derivative, and $\c
B_\mu=\varepsilon_{\mu\nu\alpha} \D^\nu B^\alpha$ is the covariant
spatial curl. The kinematic quantities are $\Theta$ (the volume
expansion of $u^\mu$-flowlines), $\omega_\mu$ (vorticity),
$\dot{u}_\mu$ (four-acceleration), and $\sigma_{\mu\nu}$ (shear).

The key equation is (\ref{7}), which is the induction equation in
covariant form. When contracted with $B^\mu$, it leads to the
conservation equation for magnetic energy density:
\begin{equation}\label{8}
\dot{\rho}_{\rm mag}+{\textstyle{4\over3}}\Theta\rho_{\rm mag}=
\sigma_{\mu\nu}\pi^{\mu\nu}\,,
\end{equation}
where
\begin{equation}
\rho_{\rm mag}={\textstyle{1\over2}}B_\mu B^\mu\,,~ \pi_{\mu\nu}=
{\textstyle{1\over3}}B^\alpha B_\alpha
h_{\mu\nu}-B_{\mu}B_{\nu}\,,
\end{equation}
are the energy density and anisotropic stress of the magnetic
field. Typically, the term on the right of equation (\ref{8}) may
be neglected, in which case $\rho_{\rm mag}$ obeys the same
evolution equation as isotropic radiation, so that
\begin{equation}\label{9}
r\equiv {\rho_{\rm mag}\over\rho_{\rm rad}}=\mbox{ constant}\,.
\end{equation}

In a Friedmann universe, where $\Theta=3H=3\dot{a}/a$ and $a$ is
the scale factor, we have from equation (\ref{8}) that
\begin{equation}
a^2B=\mbox{ constant}\,,
\end{equation}
where $B=(B_\mu B^\mu)^{1/2}$. If we choose $a=1$ at the present
time, then $a^2B$ is the comoving magnitude of the magnetic field.
Observations show that galactic and cluster fields are at the
micro-Gauss level.

Nucleosynthesis imposes limits based on the way in which a
magnetic field affects the expansion rate, the reaction rates and
the electron phase density~\cite{gr}:
\begin{equation}
a^2B\lesssim 10^{-7}\,{\rm G}\,,
\end{equation}
on cosmological scales. We can understand this limit qualitatively
by requiring that $\rho_{\rm mag}<\rho_{\rm rad}$ at
nucleosynthesis, which gives the right order of magnitude.

The upper limit from the CMB on a large-scale field is much
tighter~\cite{bfs}:
\begin{equation}
a^2B\lesssim 10^{-9}\,{\rm G}\,.
\end{equation}
This field strength corresponds to an energy density
\begin{equation}
\Omega_{\rm mag}\equiv {\rho_{\rm mag}\over \rho_{\rm crit}}\sim
10^{-5} \Omega_{\rm rad}\,,
\end{equation}
so that, roughly speaking, magnetic fields cannot induce
large-angle perturbations in the CMB above the observed level.

\section{Magnetogenesis and amplification}

{\em Protogalactic magnetogenesis,} i.e. the creation of magnetic
fields during the process of structure formation, essentially
relies upon battery-type mechanisms in which the gradients of
electron number density $n_{\rm e}$ and pressure $p_{\rm e}$ are
not aligned. Ohm's law (\ref{3}) is modified and leads to the
modified induction equation (in Newtonian form)~\cite{battery}
\begin{equation}
{\partial \vec{B}\over\partial t}=\vec{\nabla}\times(\vec{v}
\times\vec{B})+\alpha\,\vec{n}_{\rm e}\times\vec{p}_{\rm e}\,,
\end{equation}
where $\alpha$ is a constant. It follows that if the gradient
terms are non-aligned (as happens for example when shock waves
develop in collapsing clouds), then nonzero $B$ can be generated.
Very small fields are generated in this way, and typically require
strong dynamo-type amplification in order to reach the currently
observed levels.

A seed magnetic field, whether generated by battery mechanisms or
already present in the form of a primordial field, is amplified
adiabatically during gravitational collapse, simply by the fact
that field lines are frozen into the plasma, and compression of
the plasma results in compression of flux lines. This adiabatic
compression is weak, with growth roughly given by
\begin{equation}
B \propto \delta^{2/3}\,,
\end{equation}
where $\delta=\delta\rho/\rho$ is the fractional over-density of
the cloud [this neglects the shear term in equation (\ref{8})]. If
the observed galactic fields ($\sim 10^{-6}$~G) are the result
only of adiabatic compression, then the seed field required could
be up to $\sim 10^{-9}$~G (comoving). This is at the level of the
CMB limit on large-scales.

If the seed field is much weaker, then a stronger amplification is
required -- and the prime candidate mechanism for this is the
galactic dynamo~\cite{dynamo}. This is based on differential
rotation and turbulence, whereby small-scale magnetic fields are
amplified via parametric resonance. The key issue of how efficient
the dynamo is, has not been settled. There is therefore a large
uncertainty in the amount of amplification that can be achieved,
and thereby in the size of seed field that is necessary. In
general qualitative terms, the seed field will be much less than
that required for purely adiabatic compression. In terms of the
$r$-factor in equation (\ref{9}), a seed without dynamo
amplification requires $r\sim 10^{-14}$, whereas a seed with
dynamo amplification could have $r$ as low as $\sim 10^{-34}$
(this may be further reduced in the presence of a cosmological
constant~\cite{constant}).

{\em Primordial magnetogenesis} is the creation of magnetic fields
in the early universe, before the process of structure formation.
Many mechanisms have been proposed, based mainly on phase
transitions before recombination, or on inflation. In phase
transitions such as the QCD and EW transitions, local charge
separation can arise, creating local currents that can generate
(hyper-)magnetic fields~\cite{soj}. Other proposals include
bubble-wall collisions, which produce phase gradients that can
source gauge fields~\cite{kv}.

These mechanisms produce fields coherent on sub-Hubble scales. In
order to generate super-Hubble scale fields, one requires
inflationary models~\cite{inf}, or pre big bang models based on
string theory~\cite{pbb}, in which vacuum fluctuations of the
field are amplified via the dilaton. Inflation stretches
perturbations beyond the Hubble horizon and thus can in principle
generate magnetic fields on large scales. There is however a
problem in that vector perturbations are extremely small in
standard models, essentially because the vector gauge field does
not couple gravitationally to a conformally flat metric. One needs
to break conformal invariance by new high-energy couplings of the
photon (or to break gauge invariance). An example of such a
coupling is provided by the Lagrangian for scalar electrodynamics:
\begin{equation}
{\cal L}=-{\textstyle{1\over4}}F_{\mu\nu}F^{\mu\nu}-({\cal
D}_\mu\phi)^*{\cal D}^\mu\phi-V(\phi\phi^*)\,,
\end{equation}
where $\phi$ is the charged scalar field, and ${\cal
D}_\mu=\nabla_\mu-{\rm i}eA_\mu$ is the gauge-covariant
derivative, with $e$ the coupling constant.

Inflation is often followed by a preheating period in which
coherent oscillations of the inflaton produce parametric resonant
amplification of perturbations. Since the inflaton is coherent on
super-Hubble scales, this amplification can in principle affect
super-Hubble scales, without in any way violating
causality~\cite{bb}. Magnetic fields arising from inflationary
fluctuations could thus in principle be amplified via
preheating~\cite{preh}.

\section{Magnetic fields and the CMB}

In the absence of any preferred model of primordial
magnetogenesis, and in view of the complexities of
magnetohydrodynamics during structure formation (especially the
dynamo mechanism), we need cosmological observational tests for
deciding whether magnetogenesis is primordial or protogalactic. If
magnetic fields could be detected in the voids between galactic
clusters, this would be very strong evidence for a primordial
origin.

The other key observational test is provided by the CMB.
Dynamically significant magnetic fields present during
recombination must be primordial. These primordial fields have
various effects on the CMB.

In the absence of a magnetic field, the tightly coupled
baryon-photon fluid undergoes longitudinal acoustic oscillations
in density and velocity perturbations, with
\begin{equation}
\delta\,,v \propto \exp({\rm i}kc_{\rm s}\eta)\,,
\end{equation}
where $c_{\rm s}$ is the sound speed and $\eta$ is conformal time.
A magnetic field splits these modes into 3 types:\\(a) fast
magnetosonic waves, which are like sound waves, but with increased
speed,
\begin{equation}
c_{\rm s}^2~\to~c_{\rm s}^2+c_{\rm a}^2\sin^2\theta\,,
\end{equation}
where $c_{\rm a}^2=\rho_{\rm mag}/\rho$ is the Alfv\'en speed
squared and $\theta$ is the angle between $\vec{B}$ and the
propagation direction;\\ (b) slow magnetosonic waves, which have
speed $c_{\rm a}\cos\theta$ and are partly transverse in
velocity;\\ (c) incompressible Alfv\'en waves, whose speed is the
same as the slow magnetosonic waves, and for which $\delta=0$.

The fast magnetosonic waves have a direct and simple, though
small, effect on the acoustic peaks in CMB temperature
anisotropies~\cite{adgr}, based on the modification of the sound
speed. The effect of Alfv\'en modes on CMB anisotropies has also
been calculated~\cite{alf}.

Fast magnetosonic modes suffer diffusion damping just like the
non-magnetized acoustic modes. The slow magnetosonic and Alfv\'en
modes by contrast can be overdamped and survive on scales below
the Silk scale~\cite{jko}. This could play an interesting role in
structure formation.

In general, the dissipation of magnetized fluctuations injects
non-thermal energy into the photon spectrum, which introduces
chemical-potential and Compton distortions ($\mu$ and $y$
distortions) in the CMB blackbody. Upper limits on these
distortions provided by the FIRAS experiment on COBE then place
upper limits on the magnetic field strength~\cite{jko2}:
\begin{equation}
a^2B \lesssim 10^{-8}\,\mbox{G~  on scales }~ \sim0.5-600\,{\rm
kpc}\,.
\end{equation}

The anisotropic stress $\pi_{\mu\nu}$ induced by a magnetic field
can source gravitational wave perturbations during recombination.
This can be seen through the wave equation that governs the
transverse traceless magnetic part of the Weyl tensor,
$H_{\mu\nu}$, which provides a covariant description of
gravitational waves~\cite{t}:
\begin{equation}
-\ddot{H}_{\mu\nu}+\D^2H_{\mu\nu}=7H\dot{H}_{\mu\nu}+2\rho(1-w)
H_{\mu\nu}+2H\c \pi_{\mu\nu}\,,
\end{equation}
where $\c\pi_{\mu\nu}=\varepsilon_{\alpha\beta(\mu}\D^\alpha
\pi^\beta{}_{\nu)}$ is the covariant spatial tensor curl and
$w=p/\rho$. In order to keep the tensor contribution to CMB
temperature anisotropies within the observed limits, this places
upper limits on the magnetic field~\cite{dfk}.

Magnetic fields have an important effect on the polarization of
the CMB via Faraday rotation~\cite{pol}. Linearly polarized
radiation with frequency $\nu$ and wave vector $\vec{e}$, in a
magnetized plasma with free electron density $n_{\rm e}$, has its
plane of polarization rotated through an angle $\varphi$, where
\begin{equation}
{d\varphi\over dt}\propto {n_{\rm e}\over
\nu^2}\,\vec{B}\cdot\vec{e}\,.
\end{equation}
For a given line of sight $\vec{e}$, the polarization angle
$\varphi$ may be measured at different frequencies, thus providing
in principle a measure of the magnetic field strength. The Planck
experiment may be able to detect a field at the $\sim 10^{-9}$ G
level. An indirect effect of Faraday rotation is to depolarize the
CMB on small angular scales, leading to a reduction in damping and
thus a small increase in power in the temperature
anisotropies~\cite{hdz}.

Perhaps more significant than the small quantitative effects on
polarization angle and on small-scale temperature anisotropies is
an intriguing correlation introduced by magnetic fields~\cite{sf}.
Scalar perturbations can only generate E-type polarization, while
tensor perturbations generate both E- and B-type polarization. A
magnetic field also generates both E- and B-type polarization, but
in addition, it induces a correlation via the Faraday rotation
coupling in the evolution equations for polarization. This means
that the B-type polarization will be correlated with the
temperature anisotropies. Such a correlation does not arise in the
context of statistical isotropy, but a large-scale magnetic field
breaks the isotropy and produces an novel signature, which may be
more accessible to observation.

Another potentially important (although probably extremely small)
effect on CMB temperature anisotropies arises from the general
relativistic interaction between gravity and electromagnetism,
whereby electromagnetic radiation may be induced from a magnetic
field by gravitational waves~\cite{mbd}.

\section{Magnetized structure formation}

The effects of a weak cosmological magnetic field on structure
formation in the linear regime are necessarily very small. The
pioneering analysis was given in~\cite{rr} (see also~\cite{w}). In
the matter era on sub-Hubble scales, a Newtonian approach is
justified, based on the magnetized Euler equation
\begin{equation}
{\partial\vec{v}\over\partial\eta}+aH\vec{v}=-c_{\rm s}^2
\vec{\nabla}\delta-\vec{\nabla}\Phi+{1\over\rho} (\vec{\nabla}
\times \vec{B})\times \vec{B}\,,
\end{equation}
where $\Phi$ is the gravitational potential perturbation. The
standard, non-magnetized adiabatic growing mode of density
perturbations is slightly damped by magnetism~\cite{rr,tb}:
\begin{equation}
\delta \propto a^n\,,~~n={\textstyle{1\over4}} \left[-1+
5\sqrt{1-\alpha_{\rm mag} k^2}\,\right]\,,
\end{equation}
where $\alpha_{\rm mag}$ is a constant determined by $c_{\rm
a}^2$, and $k$ is the wave number. New non-adiabatic constant and
decaying modes are also introduced by the magnetic field. A
magnetic field can induce density perturbations in a homogeneous
medium, although it cannot on its own reproduce the features of
the observed power spectrum~\cite{kor}. An analysis of the complex
dynamics of magnetized damping during recombination~\cite{jko,alf}
shows that incompressible and slow magnetosonic modes may survive
on scales well below the Silk scale, and this could lead to
interesting variations on the non-magnetized scenario of structure
formation. These small-scale modes that survive damping could seed
early star or galaxy formation and could also precipitate
fragmentation of early structures.

The magnetic field also acts as a source of incompressible
rotational instabilities, which satisfy the wave
equation~\cite{tm}
\begin{eqnarray}
-\ddot{W}_\mu+\left[{c_{\rm a}^2\over 3(1+w)}\right]\D^2W_\mu &=&
(4-3w)H\dot{W}_\mu \nonumber\\&~~&{}
+{\textstyle{1\over2}}\rho\left[1-7w+3c_{\rm
s}^2(1+w)\right]W_\mu\,,
\end{eqnarray}
where $D^\mu W_\mu=0$. On small scales, these vortices may have
some interesting effects on structure formation. Magnetic fields
can generate not only vorticity, but also anisotropic distortion
in the density distribution~\cite{tm}.

On super-Hubble scales, a fully general relativistic analysis is
needed, and this is developed in~\cite{tb,tm} (see~\cite{hd} for a
dynamical-systems analysis of the equations). During the radiation
era, the non-magnetized adiabatic growing mode is incorrectly
predicted to suffer small magnetic damping via an analysis which
does not incorporate all relativistic effects. In fact, there is a
crucial {\em magneto-curvature coupling,}~\cite{tb,tm,tm2} which
arises from the non-commutation of the projected covariant
derivatives of the magnetic field:
\begin{equation}\label{}
\D_{[\mu}\D_{\nu]}B_\alpha={\textstyle{1\over2}}{\cal
R}_{\mu\nu\alpha\beta}B^\beta-\varepsilon_{\mu\nu\beta}\omega^\beta
\dot{B}_\alpha\,,
\end{equation}
where the projected curvature tensor is
\begin{equation}
{\cal R}_{\mu\nu\alpha\beta}=h_\mu{}^\sigma h_\nu{}^\chi
h_\alpha{}^\gamma h_\beta{}^\delta R_{\sigma\chi\gamma\delta} -
V_{\mu\alpha}V_{\nu\beta}+V_{\mu\beta}V_{\nu\alpha}\,,
\end{equation}
with
\begin{equation}
V_{\mu\nu}={\textstyle{1\over3}}\Theta h_{\mu\nu}+\sigma_{\mu\nu}+
\varepsilon_{\mu\nu\alpha}\omega^\alpha\,.
\end{equation}
This coupling combines with the tension of the magnetic
force-lines to reverse the damping effect and leads to a small
enhancement of the growing mode, which satisfies the
equation~\cite{tm}
\begin{equation}
a^2{d^2\delta\over da^2}-(2-c_{\rm a}^2)\delta=c_{\rm a}^2\left(
C+ 2a^2{\cal R}\right)\,,
\end{equation}
where $C$ is a constant and ${\cal R}=h^{\mu\nu}
h^{\alpha\beta}{\cal R}_{\mu\alpha\nu\beta}$ is the projected
curvature scalar.

The coupling between magnetism and curvature essentially injects
the elastic properties of magnetic field lines into space itself,
and can lead to rather unexpected dynamical and kinematical
effects~\cite{tm2,mt}.

\section{Conclusion}

Cosmic magnetic fields provide a fascinating set of unsolved
problems challenging theorists in cosmology. Not only do we need
to resolve the key question as to whether these fields are
primordial or protogalactic in origin, but we also need to develop
a satisfactory theory of magnetogenesis and amplification.
Furthermore, there are a number of open issues in calculating the
magnetic effects on structure formation and on CMB anisotropies.
The required theoretical developments will be driven by advances
in observations, both directly of magnetic fields beyond the
galactic scale, and indirectly via future advances in CMB
observations and large-scale structure surveys.

\vfill{ {\bf Acknowledgements}\\ I thank Christos Tsagas and
Alejandra Kandus for many helpful discussions. I am grateful to
the organisers of the conference for their wonderful hospitality
and kindness, with special thanks to Sayan Kar and Naresh Dadhich.
It was an honour to speak at Kharagpur, on the site of resistance
to a former colonial prison, where many freedom fighters refused
to surrender. }

\end{document}